\documentclass[preprintnumbers,nofootinbib,noshowpacs,eqsecnum,twocolumn,prd]{revtex4}

\usepackage{graphicx}
\usepackage{amsmath,amsthm,amssymb}
\usepackage{mathrsfs,bbm}
\usepackage{color,array}

\definecolor{lgray}{gray}{0.9} 		
\graphicspath{ {figures/} }

\makeatletter       

\renewcommand{\p@subsection}{}

\makeatother

\newtheorem{theorem}{Theorem}

\newcommand*{\eweakgroup}{\mbox{$SU(2)_L \times U(1)_Y$} }
\newcommand*{\emgroup}{\mbox{$U(1)_{em}$} }

\newcommand*{\unitmatrix}{\mathbbm{1}}

\newcommand*{\twomat}[1]{\underline{#1}}             
\newcommand*{\tvec}[1]{\boldsymbol{#1}}              
\newcommand*{\trans}{\mathrm{T}}                     
\newcommand*{\by}{\!\times\!}                        

\DeclareMathOperator{\trace}{tr}
\DeclareMathOperator{\mRe}{Re}
\DeclareMathOperator{\mIm}{Im}

\begin{document}

\title{Stability and symmetry breaking in the general n-Higgs-doublet model}

\author{M. Maniatis}
    \email[E-mail: ]{MManiatis@ubiobio.cl}
\affiliation{Departamento de Ciencias B\'a{}sicas, 
Universidad del B\'i{}o B\'i{}o, Casilla 447, Chill\'a{}n, Chile.}
\author{O. Nachtmann}
    \email[E-mail: ]{O.Nachtmann@thphys.uni-heidelberg.de}
\affiliation{Institut f\"ur Theoretische Physik, Philosophenweg 16, 69120
Heidelberg, Germany}


\begin{abstract}
For potentials with $n$-Higgs-boson doublets stability, electroweak symmetry breaking, 
and the stationarity equations
are discussed in detail. This is done within
the bilinear formalism which simplifies the investigation, in particular since
irrelevant gauge degrees of freedom are systematically avoided. 
For the case that the potential leads to the physically relevant 
electroweak symmetry breaking the mass matrices of the physical Higgs bosons
are given explicitly.
\end{abstract}

\maketitle


\section{Introduction}
\label{intro}
Despite the fact that the Standard Model~(SM) has only one
Higgs-boson doublet, there is no theoretical restiction
to impose a larger number of Higgs-boson doublets.
In particular, an extended Higgs sector opens
the possibility of CP violation in the Higgs potential. 
This was already shown
by T.D.~Lee for the case of the 
two-Higgs-doublet model~(THDM)~\cite{Lee:1973iz}.

Here we want to focus on the general $n$-Higgs-doublet potential,
where we assume that all doublets carry the same hypercharge.
The aim is to find precise conditions for
stability, electroweak symmetry-breaking, as well as to give
equations to find systematically all stationary points, in particular,
to detect the global minimum. It was shown that this
is indeed possible in the case of the THDM~\cite{Maniatis:2006fs} as well as in the
3HDM~\cite{Maniatis:2014oza}. Here we want to generalise these findings.
We will apply the {\em bilinear} formalism, which was 
developed in~\cite{Nagel:2004sw,Maniatis:2006fs}
and independently in~\cite{Nishi:2006tg}.
Let us note, that the one-to-one correspondance of the
gauge orbits of the Higgs-boson doublets
and the bilinears in the general nHDM was already given in~\cite{Maniatis:2006fs}.

On the experimental side there is lots of effort spent to
detect more than one physical Higgs boson for instance by the current
LHC experiments. 
On the theoretical 
side also many models have been proposed which involve an
extended Higgs sector. 
It is well known that supersymmetric models like the 
minimal and the next-to-minimal supersymmetric standard model
require extended Higgs sectors.
For reviews see for instance \cite{Nilles:1983ge}
and \cite{Maniatis:2009re,Ellwanger:2009dp}, respectively.
Two-Higgs-doublet models have been reviewed in \cite{Branco:2011iw}.
The completely general Higgs sector was considered in \cite{Bernreuther:1998rx}
in connection with possible CP-violating effects in Z-boson decays.
For further works on models with extended
Higgs sectors see for instance~\cite{Ma:2009ax, Grzadkowski:2010dj, Ferreira:2010yh, Keus:2013hya, 
Kaminska:2013mya, Muhlleitner:2014vsa, Barger:2014qva}.
Interesting relations between charge breaking, CP violating, and the 
normal vacuum in multi-Higgs-doublet models were obtained in~\cite{Barroso:2006pa}. 
Let us also mention that various aspects of the general nHDM in terms of bilinears 
have been discussed in~\cite{Maniatis:2006fs,Maniatis:2007vn,Nishi:2006tg,Nishi:2007nh,Ivanov:2010ww,Ivanov:2010wz,Ivanov:2012fp, Maniatis:2015kma}.


\section{Bilinears}
\label{bilinears}

Let us now consider the tree-level Higgs potential of models with
$n$~Higgs-boson doublets
satisfying \eweakgroup electroweak
gauge symmetry. 
The case of $n$~Higgs-boson doublets is a generalisation
of the cases with two or three doublets which were
discussed in detail in~\cite{Maniatis:2006fs,Maniatis:2014oza}.

We will assume that we have~$n\ge 2$ doublets which all 
carry the same hypercharge $y=+1/2$ and denote the
complex doublet fields by
\begin{equation}
\label{eq-doubldef}
\varphi_i(x) = 
\begin{pmatrix} 
\varphi^+_i(x) \\
 \varphi^0_i(x)
 \end{pmatrix}; \qquad i=1,\ldots,n.
\end{equation}
The most general \eweakgroup gauge invariant Higgs potential
consists solely of products of
the Higgs-boson doublets in the form
\begin{equation}
\label{eq-potterms}
\varphi_i(x)^{\dagger}\varphi_j(x),
\qquad i,j \in \{1,\ldots,n\}.
\end{equation}

We will now introduce gauge invariant bilinears, which
turn out to be convenient to discuss the properties of the Higgs potential such
as its stability and its stationary points.

To this end we introduce the $n \times 2$~matrix of the Higgs-boson fields
\begin{equation} \label{2.3}
\phi = 
\begin{pmatrix} 
\varphi^+_1 & \varphi^0_1  \\
\vdots  & \vdots\\
\varphi^+_n & \varphi^0_n
\end{pmatrix} = 
\begin{pmatrix} 
\varphi_1^\trans \\
\vdots \\
\varphi_n^\trans \\
\end{pmatrix} .
\end{equation}
All possible \eweakgroup invariant
scalar products may be arranged into the hermitian $n\by n$~matrix
\begin{gather}
\label{eq-kmat}
\twomat{K} =  \phi \phi^\dagger =
\begin{pmatrix}
  \varphi_1^{\dagger}\varphi_1 & \varphi_2^{\dagger}\varphi_1 & \ldots & \varphi_n^{\dagger}\varphi_1 \\
  \varphi_1^{\dagger}\varphi_2 & \varphi_2^{\dagger}\varphi_2 & \ldots & \varphi_n^{\dagger}\varphi_2 \\
  \vdots & & \ddots & \ \vdots\\
  \varphi_1^{\dagger}\varphi_n & \varphi_2^{\dagger}\varphi_n & \ldots & \varphi_n^{\dagger}\varphi_n  
\end{pmatrix}.
\end{gather}

A basis for the $n \times n$ matrices is given by the $n^2$ matrices
\begin{equation} \label{2.4a}
\lambda_\alpha, \quad \alpha = 0,1,\ldots,n^2-1
\end{equation}
where 
\begin{equation}\label{2.4b}
\lambda_0 = \sqrt{\frac{2}{n}} \unitmatrix_n
\end{equation}
is the conveniently scaled unit matrix and $\lambda_a$, $a=1,\ldots,n^2-1$, are the generalised Gell-Mann matrices. 
An explicit construction and numbering scheme of the generalised Gell-Mann matrices is given
in appendix~\ref{appendixA}.
We will here and in
the following assume that greek indices ($\alpha$, $\beta$, $\ldots$) run from 0 to $n^2-1$ and 
latin indices ($a$, $b$, $\ldots$) from 1 to $n^2-1$.
We find
\begin{equation} \label{2.4c}
\begin{split}
&\trace (\lambda_\alpha \lambda_\beta) = 2 \delta_{\alpha \beta},\\
&\trace (\lambda_\alpha) = \sqrt{2 n}\; \delta_{\alpha 0}.
\end{split}
\end{equation}
The decomposition of~$\twomat{K}$~\eqref{eq-kmat} reads now
\begin{equation} \label{2.5}
\twomat{K} = \frac{1}{2} K_\alpha \lambda_\alpha
\end{equation}
where the real coefficients $K_\alpha$ are given by 
\begin{equation} \label{2.6}
K_\alpha = K_\alpha^* = \trace (\twomat{K} \lambda_\alpha).
\end{equation}
Note that in particular
\begin{equation}
K_0 = \trace (\twomat{K} \lambda_0)
= \sqrt{\frac{2}{n}}\left(\varphi_1^\dagger \varphi_1+ \ldots + \varphi_n^\dagger \varphi_n \right) .
\end{equation}
With the matrix $\twomat{K}$, as defined in terms of the doublets 
in~\eqref{eq-kmat}, as well as the decomposition~\eqref{2.5}, \eqref{2.6},
we may immediately express the scalar products in terms of the bilinears.

The matrix $\twomat{K}$~\eqref{eq-kmat} is positive semidefinite
which follows directly from its definition $\twomat{K}=\phi\phi^\dagger$. 
The $n^2$ coefficients  
$K_\alpha$ of its decomposition~\eqref{2.5} are completely
fixed given the Higgs-boson fields.

The $n \times 2$~matrix $\phi$ has trivially
rank smaller or equal 2, from which follows that this holds also for the matrix~$\twomat{K}$.
As was shown in detail in~\cite{Maniatis:2006fs} (see their theorem~5),
any hermitian $n \times n$ matrix with rank equal or smaller than~2
determines the Higgs-boson fields
$\varphi_i$, $i=1,\ldots,n$ uniquely, up to a gauge transformation.

Let us now discuss the properties of the matrix~$\twomat{K}$ 
with respect to its rank.
Since the $n\times n$ matrix $\twomat{K}$ is hermitian and 
positive semidefinite we can, by a unitary
transformation~$U$, diagonalise this matrix,
 \begin{equation} \label{diagK}
 U \twomat{K} U^\dagger =
 \begin{pmatrix}
 \kappa_1 & 0 & \cdots & 0 \\
 0 & \kappa_2 & \cdots & 0 \\
 \vdots & & \ddots & \vdots\\
 0 & 0 & \cdots & \kappa_n
 \end{pmatrix},
 \end{equation}
with all $\kappa_i\ge 0$.
We define for any hermitian matrix~$\twomat{K}$ with eigenvalues 
$\kappa_1, \ldots, \kappa_n$ the symmetric sums
 \begin{equation} \label{defek}
 \begin{split}
 &s_0 := 1,\\
 &s_1 := \kappa_1+\kappa_2+ \cdots + \kappa_n,\\
 &s_2 := \sum_{1\le i_1< i_2 \le n} \kappa_{i_1} \kappa_{i_2},\\
 &s_k := \sum_{1\le i_1<i_2< \cdots < i_k \le n} \kappa_{i_1} \kappa_{i_2} \cdots \kappa_{i_k},\\
 &s_n := \kappa_1 \cdot \kappa_2 \cdot \hdots \cdot \kappa_n = \det(\twomat{K}) .
 \end{split}
\end{equation}
The hermitian matrix~$\twomat{K}$ is positive 
semidefinite if and only if
\begin{equation}\label{2.12a}
s_k \ge 0 \qquad \text{for } k=0,\ldots,n.
\end{equation}

Suppose the matrix~$\twomat{K}$ has rank~0,
then, clearly, all $\kappa_i$ have to vanish, 
corresponding to 
\begin{equation} \label{rank0}
 s_1 = s_2 = \ldots s_n =0.
 \end{equation}
Vice versa, starting with the conditions \eqref{rank0} for a hermitian matrix $\twomat{K}$,
the last condition $s_n=0$ requires that one eigenvalue has to vanish, for instance,
 $\kappa_n=0$, without loss of generality. 
The next-to-last condition in turn requires that another, say $\kappa_{n-1}=0$, and so on.
Therefore we get $\twomat{K}=0$.

Next suppose the hermitian matrix~$\twomat{K}$ has rank~$1$, then,
without loss of generality, we can assume
\begin{equation} \label{rank1}
\begin{split}
 &\kappa_1>0,\\
 &\kappa_{2} = \ldots = \kappa_n =0.
\end{split}
\end{equation}
 If follows immediately from \eqref{defek}
 \begin{equation} \label{rank1c}
 \begin{split}
  &s_1 > 0,\\
  &s_{2} = \ldots = s_n =0.
  \end{split}
 \end{equation}
On the other hand, having the conditions \eqref{rank1c} for a hermitian matrix $\twomat{K}$
fulfilled, employing~\eqref{defek},
the last condition $s_n=0$ requires that at least one $\kappa_i$ vanishes,
for instance $\kappa_n=0$ without loss of generality.
Then the next-to-last condition requires that another eigenvalue has
to vanish, for instance $\kappa_{n-1}=0$. That is, we have
$\kappa_n = \ldots = \kappa_2 =0$.
Eventually, the first condition dictates that $\kappa_1>0$, hence,
$\twomat{K}$ has rank~1 and is positive semidefinite.

Suppose the hermitian matrix~$\twomat{K}$ has rank~$2$, then,
without loss of generality, we can assume
\begin{equation} \label{rank2}
\begin{split}
 &\kappa_1>0, \kappa_2>0,\\
 &\kappa_{3} = \ldots = \kappa_n =0.
\end{split}
\end{equation}
 If follows immediately from \eqref{defek}
 \begin{equation} \label{rank2c}
 \begin{split}
  &s_1 > 0, s_2 > 0,\\
  &s_{3} = \ldots = s_n =0.
  \end{split}
 \end{equation}
On the other hand, having the conditions \eqref{rank2c} for a hermitian matrix $\twomat{K}$
fulfilled, employing~\eqref{defek},
the conditions $s_3=\ldots = s_n=0$ require that 
$\kappa_3 = \ldots = \kappa_n =0$, without loss of generality.
Then the first two conditions of~\eqref{rank2c} state $\kappa_1 + \kappa_2 >0$ and
$\kappa_1 \cdot \kappa_2>0$, that is, we have $\kappa_1>0$ and
$\kappa_2>0$. 
Hence, $\twomat{K}$ has rank~2 and is positive semidefinite.
 
 Therefore, we have shown the following theorem.
 \begin{theorem}\label{theorem} Let $\twomat{K}=K_\alpha \lambda_\alpha /2$ be a hermitian matrix.
 $\twomat{K}$ has rank~$k$ with $k=0,1,2$ and is positive semidefinite if and only if 
 \begin{equation} \label{theoremeq} 
 \begin{split}
&s_0 > 0, \ldots, s_k>0,\\
&s_{k+1} = \ldots = s_n =0.
 \end{split}
 \end{equation}
  \end{theorem}

We may express the symmetric sums $s_k$ defined in \eqref{defek}
in terms of basis-independent traces of powers of~$\twomat{K}$.
We have a recursion relation:
\begin{equation} \label{Newton}
\begin{split}
&s_0 = 1\\
&s_k = \frac{1}{k} \sum_{i=1}^k (-1)^{i-1} s_{k-i} \trace(\twomat{K}^i), \qquad k=1,\ldots,n \;.
\end{split}
\end{equation}
The derivation of~\eqref{Newton} is given in appendix~\ref{appendixA}. 
Explicitly we get for $k=1,2,3$,
\begin{equation} \label{ekexplicit}
\begin{split}
s_1= &\trace(\twomat{K}) = \sqrt{\frac{n}{2}} K_0,\\
s_2= &\frac{1}{2}  \left( \trace^2(\twomat{K}) - \trace(\twomat{K}^2) \right)\\
 = & \frac{1}{4} \left( (n-1) K_0^2 - K_a K_a \right)\\
 = & \frac{1}{4} \left(n \delta_{\alpha 0} \delta_{\beta 0} - \delta_{\alpha \beta} \right) K_\alpha K_\beta,\\
s_3= &\frac{1}{6}  \left( 
\trace^3(\twomat{K}) - 3 \trace(\twomat{K}^2) \trace(\twomat{K}) + 2 \trace(\twomat{K}^3)
	\right).
\end{split}
\end{equation}

With the theorem~\ref{theorem} and \eqref{Newton} we have expressed the rank
properties 
of the matrix $\twomat{K}$
in terms of its eigenvalues, respectively, traces of powers of the matrix~$\twomat{K}$.
  
Based on theorem~\ref{theorem}, \eqref{Newton}, and \eqref{ekexplicit},
we can show that the gauge orbits of the $n$~Higgs-boson doublet fields are in
one to one correspondance to the vectors~$\left(K_0, \ldots, K_{n^2-1}\right)^\trans$
in the $n^2$--dimensional space~$\mathbb{R}_{n^2}$ satisfying
\begin{equation} \label{2.19}
\begin{split}
&s_1\ge 0, s_2 \ge 0, \\
&s_3= \ldots =s_n=0 .
\end{split}
\end{equation}
Here the~$s_k$, $k=1,\ldots, n$, are constructed from 
the matrix~$\twomat{K}=K_\alpha \lambda_\alpha/2$
according to~\eqref{Newton}, \eqref{ekexplicit}.
That is, to every gauge orbit of the Higgs-boson fields
corresponds exactly one vector~$(K_\alpha)$ satisfying~\eqref{2.19} and vice versa.
The first two relations of~\eqref{2.19} are analogous to the {\em light cone} conditions
of the THDM; see (36) of~\cite{Maniatis:2006fs}. The remaining relations
in the case $n>2$
are specific for the nHDM.

Another way to parametrise all positive semidefinite matrices $\twomat{K}$ of
rank 1 and rank 2 is as follows.

For rank 1 the matrix $\twomat{K}$ has only one eigenvalue unequal zero, say
$\kappa_1 >0$, $\kappa_2= \ldots = \kappa_n = 0$.
Let $\tvec{w}$ be a normalised eigenvector of $\twomat{K}$ to $\kappa_1$. 
Then we have
\begin{equation} \label{2.23}
\begin{split}
&\twomat{K} = \sqrt{\frac{n}{2}} K_o \tvec{w} \tvec{w}^\dagger,\\
&\tvec{w}^\dagger \tvec{w} = 1,\\
&K_0>0, \quad \kappa_1 = \sqrt{\frac{n}{2}} K_0 .
\end{split}
\end{equation}
For the bilinears we get from \eqref{2.23}
\begin{equation} \label{2.24}
K_\alpha  = \trace (\twomat{K} \lambda_\alpha) = 
\sqrt{\frac{n}{2}} K_0 \tvec{w}^\dagger \lambda_\alpha \tvec{w} .
\end{equation}
Clearly, for any normalised vector $\tvec{w}$ from $\mathbbm{C}_n$
we get with \eqref{2.23} a positive semidefinite matrix $\twomat{K}$ of rank 1.

For rank 2 the matrix $\twomat{K}$ has exactly two eigenvalues larger than zero.
Without loss of generality we can set
\begin{equation} \label{2.25}
\begin{split}
& \kappa_1 = \sqrt{\frac{n}{2}} K_0 \sin^2 ( \chi ),\\
& \kappa_2 = \sqrt{\frac{n}{2}} K_0 \cos^2 ( \chi ),\\
&K_0 > 0, \quad 0< \chi \le \frac{\pi}{4}.
\end{split}
\end{equation}
Let $\tvec{w}_1$ and $\tvec{w}_2$ be orthonormal eigenvectors of $\twomat{K}$
to $\kappa_1$ and $\kappa_2$, respectively. We have then
\begin{equation} \label{2.26}
\begin{split}
&\twomat{K} = \sqrt{\frac{n}{2}} K_0 \bigg(
\sin^2 (\chi) \tvec{w}_1 \tvec{w}_1^\dagger 
+
\cos^2 (\chi) \tvec{w}_2 \tvec{w}_2^\dagger 
\bigg),\\
&\tvec{w}_i^\dagger \tvec{w}_j = \delta_{ij},\\
&K_0 > 0, \quad 0< \chi \le \frac{\pi}{4}.
\end{split}
\end{equation}
Here we get for the bilinears
\begin{multline} \label{2.27}
K_\alpha  = \trace (\twomat{K} \lambda_\alpha) = 
\sqrt{\frac{n}{2}} K_0 \bigg( \sin^2 ( \chi ),  \tvec{w}_1^\dagger \lambda_\alpha \tvec{w}_1 \\
+
\cos^2 ( \chi ),  \tvec{w}_2^\dagger \lambda_\alpha \tvec{w}_2 \bigg).
\end{multline}
Clearly, the reverse also holds. For any two orthonormal
vectors $\tvec{w}_1$ and $\tvec{w}_2$ the construction \eqref{2.26}
gives a positive semidefinite matrix $\twomat{K}$ of rank 2.

With \eqref{2.23} and \eqref{2.26} we have the general
parametrisation of all positive semidefinite matrices of rank 1 and rank 2,
respectively. The parametrisations of the corresponding bilinears are given in \eqref{2.24}
and \eqref{2.27}, respectively. Based on the bilinears we shall
in the following discuss the potential, 
basis transformations, stability, minimization, and electroweak symmetry breaking  
of the general $n$HDM.

\section{The \MakeLowercase{n}HDM potential and basis transformations} \label{basis}

We now write the nHDM potential 
in terms of the bilinear coefficients, $K_0$, $K_a$, $a=1,\ldots,n^2-1$,
\begin{equation}
\label{eq-vdef}
V = 
\xi_0 K_0 + \xi_a K_a + \eta_{00} K_0^2 + 2 K_0 \eta_a K_a + K_a \eta_{ab} K_b,
\end{equation}
where the  $n^2(n^2+3)/2$ parameters $\xi_0$, $\xi_a$, $\eta_{00}$, $\eta_a$
and \mbox{$\eta_{ab}=\eta_{ba}$} are real. The potential~\eqref{eq-vdef}
consists of all possible linear and
quadratic terms of the bilinears, corresponding to 
quadratic and quartic terms of the Higgs-boson doublets.
Terms of higher order should not appear in the potential with view
of renormalizability. Moreover, any constant term in the potential can be dropped
and therefore~\eqref{eq-vdef} is the most general nHDM potential.
We introduce the notation 
\begin{gather} \label{3.2}
\begin{split}
\tvec{K}=(K_1, \ldots, K_{n^2-1})^\trans, \quad
\tvec{\xi}=(\xi_1, \ldots, \xi_{n^2-1})^\trans,\\
\tvec{\eta}=(\eta_1, \ldots, \eta_{n^2-1})^\trans, \quad
E=(\eta_{ab}),\\
\left(\tilde{E}_{\alpha \beta}\right) =
\begin{pmatrix}
\eta_{00} & \eta_b\\
\eta_a & \eta_{ab}
\end{pmatrix} =
\begin{pmatrix}
\eta_{00} & \tvec{\eta}^\trans\\
\tvec{\eta} & E
\end{pmatrix}.
\end{split}
\end{gather}
With this we can write the potential~\eqref{eq-vdef} as
\begin{equation} \label{pot4}
V = \xi_\alpha K_\alpha + K_\alpha \tilde{E}_{\alpha \beta} K_\beta .
\end{equation}

Now we consider a change of basis of the Higgs-boson fields,
$\varphi_i(x) \rightarrow \varphi'_i(x)$, with
\begin{equation}
\label{eq-udef}
\begin{pmatrix} \varphi'_1(x)^\trans \\
                \vdots \\
                \varphi'_n(x)^\trans  \\                
                \end{pmatrix}
= U  \begin{pmatrix} 
\varphi_1(x)^\trans \\
\vdots \\
\varphi_n(x)^\trans
\end{pmatrix}, 
\end{equation}
where $U \in U(n)$ is a $n \times n$ unitary transformation, 
that is, $U^\dagger U = \unitmatrix_n$. From~\eqref{eq-udef} we find
$\phi'(x)=U \phi(x)$, and  
for the matrix~$\twomat{K}$~\eqref{eq-kmat} and the bilinears
\begin{equation} \label{3.2a}
\twomat{K}'(x) = U \twomat{K}(x) U^\dagger ,
\end{equation}
\begin{equation} \label{3.3}
K_0'(x) =  K_0(x), \qquad K_a'(x) = R_{ab}(U) K_b(x).
\end{equation}
Here \mbox{$R_{ab}(U)$} is defined by
\begin{equation}
U^\dagger \lambda_a U = R_{ab}(U)\,\lambda_b.
\end{equation}
The $(n^2-1) \times (n^2-1)$ matrix~\mbox{$R(U)$} has the properties
\begin{equation}
\label{eq-rprodet}
R^\ast(U)=R(U),
\quad
R^\trans(U)\, R(U) = \unitmatrix_{n^2-1},
\;
\det \left( R(U) \right) = 1,
\end{equation}
that is, $R(U)\in SO(n^2-1)$.
Let us note that the $R(U)$ form only a subset of $SO(n^2-1)$.

A pure phase transformation, $U=\exp (i \alpha) \unitmatrix_n$,
plays no role for the bilinears. We will, therefore, consider here only transformations~\eqref{eq-udef}
with $U \in SU(n)$. In the transformation of the bilinears~\eqref{3.3}
$R_{ab}(U)$ is then the $(n^2-1) \times (n^2-1)$ matrix corresponding to~$U$ in the
adjoint representation of~$SU(n)$.

Under the replacement~\eqref{3.3}, the Higgs potential~\eqref{eq-vdef} remains unchanged
if we perform an appropriate simultaneous transformation
of the parameters
\begin{equation}
\label{eq-partrafo}
\begin{alignedat}{2}
\xi'_0 &= \xi_0,  & \tvec{\xi}' &= R(U)\,\tvec{\xi}, \\
\eta'_{00} &= \eta_{00}, &  \tvec{\eta}' &= R(U)\,\tvec{\eta}, \\
 E' &= R(U)\,E\,R^\trans(U).
\end{alignedat}
\end{equation}

A realistic $n$-Higgs-doublet model contains besides the Higgs potential
kinetic terms for the Higgs-boson doublets as well as 
Yukawa couplings which couple the Higgs-boson doublets to fermions.

Under a basis transformation, that is, a transformation~\eqref{eq-udef} of the 
Higgs-boson doublets, or in terms of the bilinears, 
a transformation~\eqref{3.3}, the kinetic terms of the Higgs doublets
are kept invariant. But in general the Yukawa couplings are {\em not} invariant 
under such a change of basis.


\section{Stability of the \MakeLowercase{n}HDM}
\label{stability}

Now we study stability of the general nHDM potential~(\ref{eq-vdef}),
given in terms of the bilinears $K_0$ and~$\tvec{K}$ on the domain
determined by~\eqref{2.19}.
This is done in an analogous way to the cases with $n=2, 3$, that is the THDM
and the 3HDM; see~\cite{Maniatis:2006fs,Maniatis:2014oza}.  
The case $\sqrt{n/2} K_0=\varphi_1^\dagger \varphi_1 +
\ldots + \varphi_n^\dagger \varphi_n = 0$
corresponds to vanishing Higgs-boson fields and $V=0$.
For \mbox{$K_0 > 0$} we define
\begin{equation}
\label{eq-ksde}
\twomat{k} = \frac{\twomat{K}}{K_0}, \quad
k_\alpha = \frac{K_\alpha}{K_0}, \quad
\tvec{k} =  \left(k_1, \ldots, k_{n^2-1} \right)^\trans.
\end{equation}
Now we write the rank conditions of theorem~\ref{theorem} in terms of $\tvec{k}$.
With help of~\eqref{2.5} we see that $\twomat{K}=K_0 \cdot (\lambda_0+k_a \lambda_a)/2$.
Therefore, the expressions $s_k$~\eqref{Newton} are proportional to $K_0^k$.
We define the dimensionless expressions $\bar{s}_k$ by 
\begin{equation}
\bar{s}_k =  \frac{s_k}{K_0^k},
\end{equation}
and get from~\eqref{Newton}
\begin{equation} 
\begin{split}
\bar{s}_0 = &1,\\
\bar{s}_1= &\frac{1}{2}\trace(\lambda_0+k_a \lambda_a)=\sqrt{\frac{n}{2}},\\
\bar{s}_2= &\frac{1}{2}  \left( \frac{1}{2}\trace^2(\lambda_0+k_a \lambda_a) - 
\frac{1}{4}\trace([\lambda_0+k_a \lambda_a]^2) \right)=\\ 
& \qquad \frac{1}{4} \left( n-1-k_a k_a\right) ,\\
\bar{s}_k =& \frac{1}{k} \sum_{i=1}^k (-1)^{i-1} \bar{s}_{k-i} 
\trace\left( \left[ \frac{\lambda_0+k_a \lambda_a}{2}\right]^i\right),\, k=1,\ldots,n \;.
\end{split}
\end{equation}
In terms of the $k_a$  
we have for~$\tvec{k}$ the domain~${\cal D}_{\tvec{k}}$:
\begin{equation} \label{4.1a}
\begin{split}
&\bar{s}_2 \ge 0,\\
&\bar{s}_3=\bar{s}_4=\ldots =\bar{s}_n =0.
\end{split}
\end{equation}
The domain boundary, $\partial {\cal D}_{\tvec{k}}$, is given by 
\begin{equation} \label{4.1b}
\bar{s}_2 = \frac{1}{4} \left( n-1-k_a k_a \right) = 0 .
\end{equation}

From~(\ref{eq-vdef}) and~(\ref{eq-ksde}) we obtain, for~\mbox{$K_0 > 0$},
$V=V_2+V_4$ with
\begin{align}
\label{eq-vk}
V_2 &= K_0\, J_2(\tvec{k}),&
J_2(\tvec{k}) &:= \xi_0 + \tvec{\xi}^\trans \tvec{k},\\
\label{eq-vk4}
V_4 &= K_0^2\, J_4(\tvec{k}),&
J_4(\tvec{k}) &:= \eta_{00} 
  + 2 \tvec{\eta}^\trans \tvec{k} + \tvec{k}^\trans E \tvec{k}
\end{align}
where we introduce the functions $J_2(\tvec{k})$ and $J_4(\tvec{k})$
on the domain~\eqref{4.1a}.

Stability of the potential means that it is bounded from below.
The stability follows from the behaviour of~$V$ in the limit
\mbox{$K_0 \rightarrow \infty$}, hence, by the signs of
\mbox{$J_4(\tvec{k})$} and \mbox{$J_2(\tvec{k})$} in~\eqref{eq-vk}, \eqref{eq-vk4}.
For a model to be at least \emph{marginally} stable, the conditions
\begin{equation} \label{eq-margstab}
\begin{split}
  J_4(\tvec{k}) &> 0 \quad\text{or}\\
  J_4(\tvec{k}) &= 0 \quad\text{and}\quad  J_2(\tvec{k}) \ge 0
\end{split}
\end{equation}
for all~$\tvec{k} \in {\cal D}_{\tvec{k}}$, that is, all~$\tvec{k}$
satisfying~\eqref{4.1a}
are necessary and sufficient,
since this is equivalent to $V \ge 0$ for
$K_0 \rightarrow \infty$ in all possible allowed directions~$\tvec{k}$.
The more strict stability property $V \rightarrow \infty$ for
$K_0 \rightarrow \infty$ and any allowed ~$\tvec{k}$ 
requires $V$ to be stable either in the strong or the weak sense.
For strong stability we require
\begin{equation} \label{4.5}
  J_4(\tvec{k}) > 0
\end{equation}
for all $\tvec{k} \in {\cal D}_{\tvec{k}}$; see~\eqref{4.1a}.
For stability in the weak sense we require for all 
$\tvec{k} \in {\cal D}_{\tvec{k}}$
\begin{equation} \label{4.6}
\begin{split}
  J_4(\tvec{k}) \ge & 0,\\
  J_2(\tvec{k}) > & 0 \text{ for all } \tvec{k} \text{ where } J_4(\tvec{k}) =0.
\end{split}
\end{equation}
In order to check that $J_4(\tvec{k})$ is positive (semi-)definite, it is sufficient to
consider its value for all stationary points 
on the domain~${\cal D}_{\tvec{k}}$.
This is true because the global minimum of the continuous function \mbox{$J_4(\tvec{k})$}
is reached on the compact domain 
${\cal D}_{\tvec{k}}$, and since the global minimum is among the
stationary points.

In order to find the stationary points of~$J_4(\tvec{k})$ in the interior of
the domain ${\cal D}_{\tvec{k}}$ we 
note that here $\twomat{k}$ of \eqref{eq-ksde} has rank 2. 
Therefore, $\twomat{k}$ and $k_\alpha$ can be represented as shown in 
\eqref{2.26} and \eqref{2.27}, respectively, but setting $K_0=1$. This gives
\begin{equation} \label{4.11}
k_\alpha = \sqrt{ \frac{n}{2}} \bigg(
\sin^2 (\chi) \tvec{w}_1^\dagger \lambda_\alpha \tvec{w}_1
+
\cos^2 (\chi) \tvec{w}_2^\dagger \lambda_\alpha \tvec{w}_2
\bigg)
\end{equation}
where
\begin{equation} \label{4.12}
\begin{split}
&\tvec{w}_1^\dagger \tvec{w}_1 - 1 = 0,\\
&\tvec{w}_2^\dagger \tvec{w}_2 - 1 = 0,\\
&\frac{1}{2} \big( \tvec{w}_1^\dagger \tvec{w}_2 + \tvec{w}_2^\dagger \tvec{w}_1 \big)= 0,\\
&\frac{1}{2 i} \big( \tvec{w}_1^\dagger \tvec{w}_2 - \tvec{w}_2^\dagger \tvec{w}_1 \big) = 0,
\end{split}
\end{equation}
\begin{equation} \label{4.13}
0 < \chi \le \frac{\pi}{4}.
\end{equation}
We have to find the stationary points of
\begin{equation} \label{4.14}
J_4(\tvec{k}) = k_\alpha \tilde{E}_{\alpha \beta} k_\beta
\end{equation}
under the constraints \eqref{4.12} and \eqref{4.13}.
The variation is with respect to the real and imaginary parts of
the components  of $\tvec{w}_1$ and $\tvec{w}_2$ and to $\chi$. 
It is easy to check that the gradient matrix of the four constraints \eqref{4.12}
has rank 4. Therefore, we can use the Lagrange  method and add these
constraints with four multipliers to $J_4$ \eqref{4.14}.
We construct the function
\begin{equation} \label{4.15}
\begin{split}
&F(\tvec{w}_1^\dagger, \tvec{w}_1, \tvec{w}_2^\dagger, \tvec{w}_2, \chi, u_1, u_2, u_3, u_4) =\\
&J_4(\tvec{k}) 
- u_1(\tvec{w}_1^\dagger \tvec{w}_1-1 )
- u_2(\tvec{w}_2^\dagger \tvec{w}_2-1 )\\
&- u_3 \frac{1}{2}(\tvec{w}_1^\dagger \tvec{w}_2 + \tvec{w}_2^\dagger \tvec{w}_1)
- u_4 \frac{1}{2 i}(\tvec{w}_1^\dagger \tvec{w}_2 - \tvec{w}_2^\dagger \tvec{w}_1)
\end{split}
\end{equation}
where $\tvec{k}=(k_a)$ is to be inserted from \eqref{4.11}.
The equations for the stationary points of $J_4$ are then obtained from
\begin{equation} \label{4.16}
\begin{split}
&\nabla_{\tvec{w}_1^\dagger, \tvec{w}_2^\dagger, \chi, u_1, u_2, u_3, u_4}
F(\tvec{w}_1^\dagger, \tvec{w}_1, \tvec{w}_2^\dagger, \tvec{w}_2, \chi, u_1, u_2, u_3, u_4) = 0,\\
&\text{ for } 0<\chi < \frac{\pi}{4} .
\end{split}
\end{equation}
For the boundary value $\chi = \pi/4$ we have
\begin{multline} \label{4.17}
\nabla_{\tvec{w}_1^\dagger, \tvec{w}_2^\dagger, u_1, u_2, u_3, u_4}\\
F(\tvec{w}_1^\dagger, \tvec{w}_1, \tvec{w}_2^\dagger, \tvec{w}_2, \pi/4, u_1, u_2, u_3, u_4) = 0.
\end{multline}
The gradients of $F$ with respect to $\tvec{w}_1$ and $\tvec{w}_2$ give the hermitian
conjugate of the gradients with respect to $\tvec{w}_1^\dagger$ $\tvec{w}_2^\dagger$,
respectively, in \eqref{4.16} and \eqref{4.17}, thus, nothing new.
For $\twomat{k}$, \eqref{eq-ksde}, of rank 1 we use \eqref{2.23}, \eqref{2.24} to get
\begin{equation} \label{4.18}
\begin{split}
&\twomat{k} = \sqrt{\frac{n}{2}} \tvec{w} \tvec{w}^\dagger,\\
&k_\alpha = \sqrt{\frac{n}{2}} \tvec{w}^\dagger \lambda_\alpha \tvec{w}
\end{split}
\end{equation}
where we have the constraint
\begin{equation} \label{4.19}
\tvec{w}^\dagger \tvec{w} -1 = 0.
\end{equation}
We easily check that the gradient matrix of the constraint has here rank 1.
Therefore we add \eqref{4.19}
with one Lagrange multiplier to $J_4$ and get
\begin{multline} \label{4.20}
F(\tvec{w}^\dagger, \tvec{w}, u) = J_4(\tvec{k}) - u (\tvec{w}^\dagger \tvec{w} -1 ) =\\
\frac{n}{2} 
 \tvec{w}^\dagger \lambda_\alpha \tvec{w} \tilde{E}_{\alpha \beta}
  \tvec{w}^\dagger \lambda_\beta \tvec{w}
- u (\tvec{w}^\dagger \tvec{w} -1 ) .
\end{multline}
The equations determining the stationary points of $J_4(\tvec{k})$ 
on the boundary 
$\partial {\cal D}_{\tvec{k}}$,
that is, for $\twomat{k}$ of rank 1, are then
\begin{equation} \label{4.21}
\nabla_{\tvec{w}^\dagger, u} F(\tvec{w}^\dagger, \tvec{w}, u) =0 .
\end{equation}

All stationary points 
obtained from \eqref{4.16}, \eqref{4.17}, and \eqref{4.21}
have to fulfill the condition
$J_4(\tvec{k})>0$ for stability in the strong sense.
If for all stationary points we have $J_4(\tvec{k})\ge 0$, then 
for every solution $\tvec{k}$ with 
$J_4(\tvec{k})=0$ we have to have 
$J_2(\tvec{k})>0$ for stability in the weak sense,
or at least
$J_2(\tvec{k})=0$ for {\em marginal} stability.
If none of these conditions is fulfilled, that 
is, if we find at least one stationary direction~$\tvec{k}$ 
with $J_4(\tvec{k})<0$ or $J_4(\tvec{k})=0$ but 
$J_2(\tvec{k})<0$, the
potential is unstable.

Our discussion above of the stability conditions for the nHDM
potential generalises the results for the THDM and the 3HDM in \cite{Maniatis:2006fs}
and \cite{Maniatis:2014oza}, respectively.
We have been careful to use in our present paper a compatible notation.
The stability properties of the general THDM and 3HDM potentials were analysed in detail in 
chapters 4 of \cite{Maniatis:2006fs} and \cite{Maniatis:2014oza}, respectively.
Also explicit examples of THDM and 3HDM potentials, using conventional parametrisations,
were discussed in these references.


\section{Electroweak symmetry breaking in the \MakeLowercase{n}HDM}

Now we assume that the nHDM potential is stable, that is, it is bounded from below.
This means that the global minimum will be among the stationary points of~$V$.
We now want to distinguish the different cases of minima 
with respect to the underlying electroweak symmetry. We shall in the following
present the corresponding stationarity equations.

We have seen in section~\ref{bilinears}, 
that the space of the Higgs-boson doublets is determined, up to electroweak gauge transformations,
by the space of the hermitian $n \times n$ matrices $\twomat{K}$ with 
rank smaller or equal~2. 
Based on the fact that the rank of the matrix  $\twomat{K}$
is equal to the rank of the Higgs-boson field matrix~$\phi$~\eqref{2.3}
we can distinguish the different types of minima with respect to
electroweak symmetry breaking as follows.
We start with writing at the global minimum, that is, the vacuum configuration,
the $n \times 2$~matrix of the Higgs-boson fields as
\begin{equation}
\label{5.1}
\langle \phi  \rangle= 
\begin{pmatrix} 
v^+_1 & v^0_1  \\
\vdots & \vdots\\
v^+_n & v^0_n  \\
\end{pmatrix}.
\end{equation}
Suppose, this matrix has rank 2, then we cannot, by a \eweakgroup 
transformation, get a form with all 
charged components~$v^+_i$, $i=1,\ldots,n$ vanishing.
Hence, the \eweakgroup group is fully broken. 
Next, suppose that at the global minimum the matrix 
$\langle \phi \rangle$ has rank one. Then we can, by a
\eweakgroup transformation get a form with all charged 
components~$v_i^+$ vanishing. Further, we can identity the unbroken~$U(1)$ 
gauge group with the electromagnetic gauge group~\emgroup. 
Hence, a minimum with rank one corresponds to the electroweak-symmetry
breaking \eweakgroup~$\rightarrow$~\emgroup\!\!.
Eventually, suppose we get a vanishing matrix at the minimum, $\langle \phi  \rangle =0$.
This corresponds to an unbroken electroweak symmetry.
Let us note that only a minimum with a partially broken electroweak symmetry
is physically acceptable.

We study now the matrix~$\twomat{K}_v$ corresponding 
to $\langle \phi  \rangle$~\eqref{5.1}
\begin{equation} \label{5.2}
\twomat{K}_v = \langle \phi \rangle \langle \phi \rangle^\dagger = 
\frac{1}{2} K_{v \alpha} \lambda_\alpha .
\end{equation}
For an acceptable vacuum~$\langle \phi  \rangle$, $\twomat{K}_v$
must have rank~1. From theorem~\ref{theorem} we see that $\twomat{K}_v$
has rank~1 and is positive semidefinite if and only if
\begin{equation} \label{5.3}
\begin{split}
&\trace \twomat{K}_v = \sqrt{\frac{n}{2}} K_{v 0} > 0,\\
& \langle s_2 \rangle = \ldots = \langle s_n \rangle = 0.
\end{split}
\end{equation}
We can bring the vacuum value~$\langle \phi \rangle$ of rank~1,
by suitable~\eweakgroup  and $U(n)$ transformations~\eqref{eq-udef}, to the form
\begin{equation} \label{5.4}
\langle \phi \rangle =
\begin{pmatrix}
0 & 0 \\
\vdots & \vdots\\
0 & v_0/\sqrt{2}
\end{pmatrix}, \qquad v_0 >0.
\end{equation}
In a realistic model $v_0$ must be the usual Higgs-boson
vacuum expectation value,
\begin{equation} \label{5.5}
v_0 \approx 246~\text{GeV}.
\end{equation}
With~\eqref{5.4} we find in this basis a simple form for
$\twomat{K}_v$ respectively $K_{v \alpha}$:
\begin{equation} \label{5.6}
\begin{split}
\twomat{K}_v = &
\frac{1}{2}
\begin{pmatrix}
0 & \hdots & 0 & 0 \\
\vdots & \ddots & \vdots & \vdots \\
0 & \hdots & 0 & 0 \\
0 & \cdots & 0 & v_0^2
\end{pmatrix} 
= \frac{1}{2} K_{v \alpha} \lambda_\alpha,\\
\big( K_{v \alpha} \big) = &
\frac{1}{\sqrt{2n}} v_0^2 \left(1,0, \hdots, 0, -\sqrt{n-1} \right)^\trans . 
\end{split}
\end{equation}

We note that another possible choice for the vacuum expectation value,
achievable by suitable transformations from \eweakgroup and $U(n)$ \eqref{eq-udef}, is
\begin{equation} \label{5.6a}
\langle \phi \rangle =
\begin{pmatrix}
0 & v_0/\sqrt{2} \\
\vdots & \vdots \\
0 & 0
\end{pmatrix}, \qquad v_0 >0.
\end{equation}
Here we get
\begin{equation} \label{5.6b}
\begin{split}
\twomat{K}_v = &
\frac{1}{2}
\begin{pmatrix}
v_0^2  & 0 & \cdots & 0 \\
0      & 0 & \cdots & 0 \\
\vdots & \vdots & \ddots & \vdots\\
0      & 0 & \cdots & 0
\end{pmatrix}.
\end{split}
\end{equation}
In the cases where~$\langle \phi \rangle$ of \eqref{5.1} has rank~2
or rank~0 also the matrix~$\twomat{K}_v$, \eqref{5.2}, has rank~2 or zero,
respectively. The corresponding conditions for~$\twomat{K}_v$ are given explicitly
in theorem~\ref{theorem} if we replace all expressions by their vacuum expectation values,
that is, $\twomat{K}$ by $\twomat{K}_v$, $K_\alpha$ by $K_{v \alpha}$ and
$s_i$ by $\langle s_i \rangle$. We summarise our findings
for the vacuum expectation values to a given potential~$V$ as follows.

Suppose~$\langle \phi \rangle$ is the vacuum expectation value of the Higgs-boson field
matrix to a given, stable, potential~$V$ and $\twomat{K}_v = \langle \phi \rangle 
\langle \phi \rangle^\dagger = K_{v \alpha} \lambda_\alpha / 2$.
The gauge symmetry \eweakgroup is fully broken by the vacuum if and only if
\begin{equation} \label{5.7}
K_{v 0}>0, \qquad  (n-1) K_{v 0}^2 - K_{v a} K_{v a} >0.
\end{equation}
We have the breaking \eweakgroup $\to$ \emgroup if and only if
\begin{equation} \label{5.8}
K_{v 0}>0, \qquad  (n-1) K_{v 0}^2 - K_{v a} K_{v a} =0.
\end{equation}
We have no breaking of \eweakgroup if and only if
\begin{equation} \label{5.9}
K_{v \alpha} =0.
\end{equation}
Clearly, we have always
\begin{equation} \label{5.10}
\langle s_3 \rangle = \cdots = \langle s_n \rangle = 0 .
\end{equation}


\section{Stationary points}
\label{stationarity}

Now suppose we have a stable potential. We shall study
the stationarity equations with view on the electroweak symmetry
breaking behavior.
If the potential is stable, the global minimum 
is among the stationary points of~$V$.
We classify the stationary points by the rank of the stationarity matrix~$\twomat{K}$.
We will apply the conditions for~$\twomat{K}$ having 
rank~0, 1, 2  as given in theorem~\ref{theorem} and \eqref{2.23} to \eqref{2.27}.

Rank~0, that is, $\twomat{K}=0$, respectively $K_\alpha=0$, $\alpha=0,\ldots,n^2-1$,
corresponds to a stationary point of~$V$ with value~$V(K_\alpha)=0$.

All stationarity matrices $\twomat{K} = K_\alpha \lambda_\alpha /2$
of rank~1 are obtained from the following system of equations.
We represent $\twomat{K}$ of rank 1 according to \eqref{2.23}. Then $K_\alpha$
is given by \eqref{2.24} and $V$ \eqref{pot4} by
\begin{equation} \label{6.1}
\begin{split}
V(K_\alpha)& = \xi_\alpha K_\alpha + K_\alpha \tilde{E}_{\alpha \beta} K_\beta \\
& = \xi_\alpha \sqrt{\frac{n}{2}} K_0 \tvec{w}^\dagger \lambda_\alpha \tvec{w} \\
& +
\big(\sqrt{\frac{n}{2}} K_0\big)^2 (\tvec{w}^\dagger \lambda_\alpha \tvec{w})
\tilde{E}_{\alpha \beta} (\tvec{w}^\dagger \lambda_\beta \tvec{w})
\end{split}
\end{equation}
where
\begin{equation} \label{6.2}
K_0>0, \qquad \tvec{w}^\dagger \tvec{w} -1 = 0.
\end{equation}
Taking the constraint equation in \eqref{6.2} into account
with a Lagrange multiplier $u$ we get the following function to determine
the stationary points of $V$ with $\twomat{K}$ of rank 1
\begin{equation} \label{6.3}
F(\tvec{w}^\dagger, \tvec{w}, K_0, u) = V(K_\alpha) - u (\tvec{w}^\dagger \tvec{w} -1 ).
\end{equation}
The gradient matrix of the constraint has rank 1 as required and we get the equations
\begin{equation} \label{6.4}
\begin{split}
&\nabla_{\tvec{w}^\dagger, K_0, u} F(\tvec{w}^\dagger, \tvec{w}, K_0, u) =0,\\
&K_0 > 0.
\end{split}
\end{equation}

All stationarity matrices $\twomat{K}= K_\alpha \lambda_\alpha /2$ of rank~2 are
obtained from the following system of equations.
We represent $\twomat{K}$ of rank 2 and the corresponding $K_\alpha$
as in \eqref{2.26} and \eqref{2.27}, respectively, and take
the constraints for $\tvec{w}_i$ into account with the help of four
Lagrange multipliers; cf. \eqref{4.12}, \eqref{4.15}.
We have then to determine the stationary points of the function
\begin{equation} \label{6.5}
\begin{split}
&F(\tvec{w}_1^\dagger, \tvec{w}_1, \tvec{w}_2^\dagger, \tvec{w}_2, \chi, K_0, u_1, u_2, u_3, u_4) =\\
&
V(K_\alpha) 
- u_1(\tvec{w}_1^\dagger \tvec{w}_1-1 )
- u_2(\tvec{w}_2^\dagger \tvec{w}_2-1 )\\
&
- u_3 \frac{1}{2}(\tvec{w}_1^\dagger \tvec{w}_2 + \tvec{w}_2^\dagger \tvec{w}_1)
- u_4 \frac{1}{2 i}(\tvec{w}_1^\dagger \tvec{w}_2 - \tvec{w}_2^\dagger \tvec{w}_1).
\end{split}
\end{equation}
The stationarity equations are then
\begin{multline} \label{6.6}
\nabla_{\tvec{w}_1^\dagger, \tvec{w}_2^\dagger, \chi, K_0, u_1, u_2, u_3, u_4}\\
F(\tvec{w}_1^\dagger, \tvec{w}_1, \tvec{w}_2^\dagger, \tvec{w}_2, \chi, K_0, u_1, u_2, u_3, u_4) = 0,\\
 0< \chi < \frac{\pi}{4}, \quad K_0>0. 
\end{multline}
For $\chi= \pi/4$ we get
\begin{multline} \label{6.7}
\nabla_{\tvec{w}_1^\dagger, \tvec{w}_2^\dagger, K_0, u_1, u_2, u_3, u_4}\\
F(\tvec{w}_1^\dagger, \tvec{w}_1, \tvec{w}_2^\dagger, \tvec{w}_2, \pi/4, K_0, u_1, u_2, u_3, u_4) = 0,\\
 K_0>0. 
\end{multline}

The stationarity matrix~$\twomat{K}=K_\alpha \lambda_\alpha /2$ with the
lowest value of $V(K_0,\ldots,K_{n^2-1})$ gives the global-minimum matrix
$\twomat{K}_v$ of the potential.
In general there may 
be degenerate global minima with the same potential value.
It was shown that systems of equations of the
type~\eqref{6.4}, \eqref{6.6}, and \eqref{6.7} can be solved via
the Groebner-basis approach or homotopy continuation;
see for instance~\cite{Maniatis:2006jd, Maniatis:2012ex}.


\section{The potential after symmetry breaking}

In this section we present the calculation of the physical Higgs-boson
masses in the nHDM.
Suppose that the potential is stable and leads to the desired electroweak
symmetry breaking, that is, $\twomat{K}_v$ has rank 1. 
From the previous discussion follows that the global minimum has then to be
obtained from a solution of the set of 
equations~\eqref{6.4}. 

Using~\eqref{6.1} we can write \eqref{6.4} explicitly as follows
\begin{multline} \label{7.1}
\sqrt{\frac{n}{2}} K_0 \bigg[ 
\xi_\alpha + 2 \tilde{E}_{\alpha \beta} \sqrt{\frac{n}{2}} K_0 
(\tvec{w}^\dagger \lambda_\beta \tvec{w}) \bigg]
\lambda_\alpha \tvec{w}\\
 - u \tvec{w} = 0,
\end{multline}
\begin{equation} \label{7.2}
\tvec{w}^\dagger \tvec{w} - 1 = 0,
\end{equation}
\begin{equation} \label{7.3}
\bigg[ 
\xi_\alpha + 2 \tilde{E}_{\alpha \beta} \sqrt{\frac{n}{2}} K_0 
(\tvec{w}^\dagger \lambda_\beta \tvec{w}) \bigg]
(\tvec{w}^\dagger \lambda_\alpha \tvec{w}) = 0,
\end{equation}
\begin{equation}
K_0>0 .
\end{equation}
Multiplying \eqref{7.1} with $\tvec{w}^\dagger$ from left and using
\eqref{7.2} and \eqref{7.3} we find
\begin{equation} \label{7.4}
u = 0.
\end{equation}

The vacuum value~$\twomat{K}_v$ is solution of this system of equations.
In the following we will always work in a basis where $\langle \phi \rangle$
and $\twomat{K}_v$ have the forms~\eqref{5.4} and \eqref{5.6}, respectively.
Furthermore, it is convenient to use instead of $\alpha = 0, 1, \ldots, n^2-2, n^2-1$ 
the basis $+,1,\ldots,n^2-2,-$; see appendix \ref{appendixA}.
Thus, all indices $\rho$, $\sigma$, ... run over this latter index set in the following.
From \eqref{5.6} we find
\begin{equation} \label{7.4a}
\begin{split}
&\twomat{K}_v = \frac{1}{2} v_0^2 \tvec{e}_n \tvec{e}_n^\dagger = \frac{1}{2\sqrt{2}} v_0^2 \lambda_-,\\
& \left(K_{v \rho} \right) = \begin{pmatrix} 0,& \ldots,& 0,& \frac{1}{\sqrt{2}} v_0^2 \end{pmatrix},\\
&K_{v -} = \frac{1}{\sqrt{2}} v_0^2.
\end{split}
\end{equation}
Here and in the following $\tvec{e}_l$, $l \in \{1,\ldots,n\}$, are the usual
Cartesian unit vectors in $\mathbbm{C}_n$. We get now that for the solution
vector $\tvec{w}$ in \eqref{7.1} we have
\begin{equation} \label{7.4b}
\tvec{w} = \tvec{e}_n
\end{equation}
and that
\begin{equation} \label{7.4c}
\sqrt{\frac{n}{2}} K_0 \tvec{w}^\dagger  \lambda_\rho \tvec{w} = K_{v \rho} .
\end{equation} 
We define
\begin{equation} \label{7.4d}
\begin{split}
\zeta_\rho &= \xi_\rho + 2 \tilde{E}_{\rho \sigma} K_{v \sigma}\\
&=  \xi_\rho + 2 \tilde{E}_{\rho -} K_{v -}.
\end{split}
\end{equation}
With this we can write \eqref{7.1}, using $K_0>0$, in the basis
$+,1,\ldots,n^2-2,-$ as
\begin{equation} \label{7.4e}
\zeta_\rho \lambda_\rho \tvec{e}_n = 0.
\end{equation}
From the explicit construction and numbering scheme of the matrices $\lambda_\rho$ in
appendix \ref{appendixA} we see that we have
\begin{equation} \label{7.4f}
\begin{split}
&\lambda_\rho \tvec{e}_n = 0, \qquad \text{for } \rho = +,1,\ldots,(n-1)^2-1 ,\\
& \lambda_{(n-1)^2} \tvec{e}_n = \tvec{e}_1,\\
& \lambda_{(n-1)^2+1} \tvec{e}_n = - i \tvec{e}_1,\\
& \qquad \vdots\\
& \lambda_{n^2-3} \tvec{e}_n = \tvec{e}_{n-1},\\
& \lambda_{n^2-2} \tvec{e}_n = -i \tvec{e}_{n-1},\\
& \lambda_{-} \tvec{e}_n = \sqrt{2} \tvec{e}_{n} .
\end{split}
\end{equation}
Therefore, \eqref{7.4e} gives
\begin{multline} \label{7.4g}
\big( \zeta_{(n-1)^2} - i \zeta_{(n-1)^2+1} \big) \tvec{e}_1 + \ldots +\\
\big( \zeta_{n^2-3} - i \zeta_{n^2-2} \big) \tvec{e}_{n-1}\\
+
\zeta_- \sqrt{2} \tvec{e}_n = 0.
\end{multline}
Since all $\zeta_\rho$ are real we get as a result of the stationarity equations
for $\twomat{K}$ of rank 1 in our basis
\begin{multline} \label{7.4h}
\zeta_\rho =0 \\ \text{for } \rho = (n-1)^2, (n-1)^2+1, \ldots, (n^2-2), - .
\end{multline}

Now we turn to the Higgs-field matrix~$\phi$. As stated above we work
in the basis where, in the unitary gauge, the vacuum-expectation value
$\langle \phi \rangle$ has the form~\eqref{5.4}.
For the original Higgs fields expressed in terms of the physical fields we get then
\begin{equation} \label{scunitary}
\begin{split}
\varphi_{i}(x) &= 
  \begin{pmatrix} H_{i}^+(x)\\ \frac{1}{\sqrt{2}} \left( H_{i}^0(x)+ i A_{i}^0(x) \right)  \end{pmatrix},
  \qquad  i=1, \ldots, n-1
\\
\varphi_n(x) &= \frac{1}{\sqrt{2}}
  \begin{pmatrix} 0\\ v_0 + h_0(x) \end{pmatrix},
\end{split}
\end{equation}
with $v_0$ real and positive, neutral fields $h_0(x)$, $H_i^0(x)$, $A_i^0(x)$, 
as well as the complex charged fields $H_i^+(x)$ with $i=1,\ldots,n-1$. 
The negatively charged
Higgs-boson fields are defined by $H_{i}^-(x) = \left(H_{i}^+(x)\right)^\dagger$.
Hence, we get in the nHDM the physical fields
\begin{equation} \label{fields-ph}
\begin{split}
&\text{$2n-1$ neutral fields:} \quad  H_i^0(x), A_i^0(x), h_0(x),\\
&\text{$n-1$ charged fields:} \quad H_i^+(x),
\end{split}
\end{equation}
with $\quad i=1,\ldots,n-1$.
It is clear that the $n$ original complex doublets of any nHDM, corresponding to $4n$ real degrees of freedom,
yield $2n-1$ real fields and $n-1$ complex fields, with the 3 remaining degrees of freedom
absorbed via the mechanism of electroweak symmetry breaking. 
Expressing the bilinears in the parametrization~\eqref{scunitary}
via \eqref{eq-kmat} and \eqref{2.5} we can write
the potential in terms of
the physical fields~\eqref{fields-ph}.
We start by expanding all quantities in powers of the physical fields.
This gives for the field matrix
\begin{equation}\label{7.10}
\begin{split}
&\phi(x) = \langle \phi \rangle + \phi^{(1)}(x),\\
&\phi^{(1)}(x) = 
\begin{pmatrix}
H_1^+(x) & \frac{1}{\sqrt{2}} \left( H_1^0(x) + i A_1^0(x) \right)\\
\vdots & \vdots\\
H_{n-1}^+(x) & \frac{1}{\sqrt{2}} \left( H_{n-1}^0(x) + i A_{n-1}^0(x) \right)\\
0 & \frac{1}{\sqrt{2}} h_0(x)
\end{pmatrix}.
\end{split}
\end{equation}
For~$\twomat{K}(x)$ we get
\begin{equation} \label{7.11}
\begin{split}
&\twomat{K}(x) = \twomat{K}_v + \twomat{K}^{(1)}(x) + \twomat{K}^{(2)}(x),\\
&\twomat{K}^{(1)}(x) = \phi^{(1)}(x) \langle \phi \rangle^\dagger + \langle \phi \rangle 
\phi^{(1)\dagger}(x),\\
&\twomat{K}^{(2)}(x) = \phi^{(1)}(x) \phi^{(1)\dagger}(x).
\end{split}
\end{equation}
Explicitly we get for $K^{(1)}_+(x)$
\begin{equation}\label{7.11a}
\begin{split}
K^{(1)}_+(x) &= \trace \left(\lambda_+ \twomat{K}^{(1)}(x) \right)\\
&=  \trace \left( \langle \phi \rangle^\dagger \lambda_+ \phi^{(1)}(x) + h.c.\right)\\
&= 0,
\end{split}
\end{equation}
Similarly we show that the only non-zero components of~$K^{(1)}_\rho(x)$
are as follows:
\begin{equation} \label{7.11c}
\begin{split}
K^{(1)}_\rho(x) = & \; v_0 H_l^0(x),\\ 
&\text{for } \rho = (n-1)^2+2l-2,\\
K^{(1)}_\rho(x) = & -v_0 A_l^0(x),\\ 
&\text{for } \rho = (n-1)^2+2l-1,\\
K^{(1)}_-(x) = & v_0 \sqrt{2} h_0(x)\\ 
\end{split}
\end{equation}
where  $l=1,\ldots,n-1$.

For the potential we write
\begin{equation} \label{7.12}
V = V^{(0)} + V^{(1)} + V^{(2)} + V^{(3)} + V^{(4)}
\end{equation}
and with~\eqref{pot4} and \eqref{7.11} we get
\begin{equation} \label{7.13}
\begin{split}
V^{(0)} = & K_{v \rho} \xi_\rho + K_{v \rho} \tilde{E}_{\rho \sigma} K_{v \sigma},\\
V^{(1)} = & K_{\rho}^{(1)}(x) \xi_\rho + 2 K_{\rho}^{(1)}(x) \tilde{E}_{\rho \sigma} K_{v \sigma},\\
V^{(2)} = & K_{\rho}^{(2)}(x) \xi_\rho + 2 K_{\rho}^{(2)}(x) \tilde{E}_{\rho \sigma} K_{v \sigma}\\
&+ K_{\rho}^{(1)}(x) \tilde{E}_{\rho \sigma} K_{\sigma}^{(1)}(x),\\
V^{(3)} = & 2 K_{\rho}^{(2)}(x) \tilde{E}_{\rho \sigma} K_{\sigma}^{(1)}(x),\\
V^{(4)} = & K_{\rho}^{(2)}(x) \tilde{E}_{\rho \sigma} K_{\sigma}^{(2)}(x).
\end{split}
\end{equation}
We shall now simplify the expressions for $V^{(0)}$, $V^{(1)}$, and $V^{(2)}$
using~\eqref{7.4a}, \eqref{7.4h}, and \eqref{7.11c}. 
Writing~$V^{(0)}$ as 
\begin{gather} \label{7.14}
\begin{split}
V^{(0)} &=  \frac{1}{2} K_{v \rho} \xi_\rho + 
\frac{1}{2} K_{v \rho} \big[\xi_\rho + 2 \tilde{E}_{\rho \sigma} K_{v \sigma} \big]\\
&=  \frac{1}{2} K_{v \rho} \big( \xi_\rho + \zeta_\rho \big)
\end{split}
\end{gather}
we find with~\eqref{7.4a} and \eqref{7.4h}
\begin{multline} \label{7.15}
V^{(0)} = \frac{1}{2} K_{v \rho} \xi_\rho = \frac{1}{2} K_{v-} \xi_- \\
= \frac{1}{2} v_0^2 \frac{1}{\sqrt{2n}}
\left( \xi_0 - \sqrt{n-1} \xi_{n^2-1} \right)
\end{multline}
which is the potential value at the vacuum.
Next we consider~$V^{(1)}$. With \eqref{7.4d}, \eqref{7.4h}, and \eqref{7.11c} we get
\begin{equation} \label{7.16}
\begin{split}
V^{(1)} = & K_{\rho}^{(1)}(x) \bigg[ \xi_\rho + 2 \tilde{E}_{\rho \sigma} K_{v \sigma} \bigg]\\
 = & K_\rho^{(1)}(x) \zeta_\rho = 0  
\end{split}
\end{equation}
since for each term in the above sum either
$K_{\rho}^{(1)}(x)=0$ or $\zeta_\rho=0$, $\rho = +,1,\ldots, n^2-2,-$.
This result must come out since we are expanding around the true minimum of the potential.

Finally we consider~$V^{(2)}$. Using again~\eqref{7.4d}, \eqref{7.4h}, 
\eqref{7.11}, and \eqref{7.13} we can write this as
\begin{equation} \label{7.17}
\begin{split}
V^{(2)} = &  K_{\rho}^{(2)}(x)  \zeta_\rho + K_{\rho}^{(1)}(x) \tilde{E}_{\rho \sigma} K_{\sigma}^{(1)}(x) \\
 = & \trace \left(\phi^{(1) \dagger}(x)  \zeta_\rho \lambda_\rho \phi^{(1)}(x) \right) \\
 & +  K_{\rho}^{(1)}(x) \tilde{E}_{\rho \sigma} K_{\sigma}^{(1)}(x).
\end{split}
\end{equation}
Here in the first term the sum runs only over $\rho = +,1,\ldots, (n-1)^2-1$ due to
\eqref{7.4h}, in the second term only over $\rho, \sigma = (n-1)^2, \ldots, n^2-2,-$
due to \eqref{7.11a} and \eqref{7.11c}.
The evaluation of these terms is straightforward. 
We define the fields
\begin{equation} \label{7.18}
\begin{split}
{\cal H}^{+}(x) = & \left( H_1^{+}(x), \ldots, H_{n-1}^{+}(x) \right)^\trans,\\
{\cal H}^{0}(x) = & \left( H_1^{0}(x), \ldots, H_{n-1}^{0}(x) \right)^\trans,\\
{\cal A}^{0}(x) = & \left( A_1^{0}(x), \ldots, A_{n-1}^{0}(x) \right)^\trans.
\end{split}
\end{equation}
Furthermore, we define $(n-1)\times (n-1)$ matrices
\begin{equation} \label{7.19}
{\mathscr M}_{\text ch}^2 = \bigg( \zeta_+ \lambda_+  +
\sum_{\rho=1}^{(n-1)^2-1}  \zeta_\rho \lambda_\rho \bigg)
\bigg|_{\begin{matrix}{\small\text{restricted to}}\\{\small\text{the first $n-1$ dimensions}}\end{matrix}},
\end{equation}
\begin{equation} \label{7.20}
\begin{split}
&{\cal E}_{HH} = (2 v_0^2 \tilde{E}_{(n-1)^2+2l-2, (n-1)^2+2l'-2}),\\
&{\cal E}_{HA} = (-2 v_0^2 \tilde{E}_{(n-1)^2+2l-2, (n-1)^2+2l'-1}),\\
&{\cal E}_{AH} = (-2 v_0^2 \tilde{E}_{(n-1)^2+2l-1, (n-1)^2+2l'-2}),\\
&{\cal E}_{AA} = (2 v_0^2 \tilde{E}_{(n-1)^2+2l-1, (n-1)^2+2l'-1})\\
& \quad \text{where}\\
& \quad l,l' \in \{1, \ldots, n-1 \}.
\end{split}
\end{equation}
We also need the $(n-1)\times 1$ matrices 
\begin{equation} \label{7.22}
\begin{split}
{\cal E}_{H-}& = (2 \sqrt{2} v_0^2 \tilde{E}_{(n-1)^2+2l-2, -})\\
& = (-2 \xi_{(n-1)^2+2l-2}),\\
{\cal E}_{A-}& = (-2 \sqrt{2} v_0^2 \tilde{E}_{(n-1)^2+2l-1, -})\\
& = (2 \xi_{(n-1)^2+2l-1}),
\end{split}
\end{equation}
the $1\times (n-1)$ matrices
\begin{equation} \label{7.23}
\begin{split}
{\cal E}_{-H}& = (2 \sqrt{2} v_0^2 \tilde{E}_{-, (n-1)^2+2l'-2})\\
& = (-2 \xi_{(n-1)^2+2l'-2}),\\
{\cal E}_{-A}& = (-2 \sqrt{2} v_0^2 \tilde{E}_{-,(n-1)^2+2l'-1})\\
& = (2 \xi_{(n-1)^2+2l'-1})
\end{split}
\end{equation}
and the scalar
\begin{equation} \label{7.24}
\begin{split}
{\cal E}_{--}& = 4 v_0^2 \tilde{E}_{--} = -2 \sqrt{2} \xi_-\\
& = - \frac{4}{\sqrt{2n}} \big( \xi_0 - \sqrt{n-1} \xi_{n^2-1} \big)\\
& = - \frac{8}{v_0^2} V^{(0)}.
\end{split}
\end{equation}
In \eqref{7.22} to \eqref{7.24} we used \eqref{7.4h} and \eqref{7.15}.
With all this we obtain for $V^{(2)}$
\begin{equation} \label{7.25}
\begin{split}
V^{(2)} & = {\cal H}^{+ \dagger}(x)\; {\mathscr M}_{\text ch}^2\;  {\cal H}^{+}(x)\\
& + 
\begin{pmatrix} {\cal H}^{0 \trans}(x),& {\cal A}^{0 \trans}(x),& h_0(x) \end{pmatrix}
\frac{1}{2} {\mathscr M}_{\text n}^2 
\begin{pmatrix} {\cal H}^{0}(x),\\ {\cal A}^{0}(x),\\ h_0(x) \end{pmatrix}
\end{split}
\end{equation}
where the mass matrix squared of the charged fields, ${\mathscr M}_{\text ch}^2$, is
given in \eqref{7.19} and
that of the neutral fields, ${\mathscr M}_{\text n}^2$, is given by
\begin{equation} \label{7.26}
{\mathscr M}_{\text n}^2 =
\begin{pmatrix}
\mRe ({\mathscr M}_{ch}^2) + {\cal E}_{HH} & -\mIm ({\mathscr M}_{ch}^2) + {\cal E}_{HA} & {\cal E}_{H-}\\
\mIm ({\mathscr M}_{ch}^2) + {\cal E}_{AH} & \mRe ({\mathscr M}_{ch}^2) + {\cal E}_{AA} & {\cal E}_{A-}\\
{\cal E}_{-H} & {\cal E}_{-A}& {\cal E}_{--}
\end{pmatrix}.
\end{equation}

Since we have assumed that we are dealing with the true vacuum,~$V^{(0)}$ \eqref{7.15} must
be below or at most equal to the potential value at $\twomat{K}=0$. That is, we must have
\begin{equation} \label{7.27}
V^{(0)} \le 0
\end{equation}
which implies, from~\eqref{7.15},
\begin{equation} \label{7.28}
\xi_0 - \sqrt{n-1} \xi_{n^2-1} \le 0 .
\end{equation}
Usually the true vacuum is required to be below the value $V=0$ 
corresponding to the trivial stationary point $K_\alpha=0$ and then
the strict inequalities must hold in \eqref{7.27} and \eqref{7.28}.
For the true vacuum the squared mass matrices of the physical Higgs bosons
must be positive semidefinite:
\begin{equation} \label{7.29}
{\mathscr M}_{\text ch}^2 \ge 0, \quad 
{\mathscr M}_{\text n}^2 \ge 0.
\end{equation}
Looking at ${\mathscr M}_{ch}^2$ we see that in general it will not
lead to mass degeneracy of all charged physical Higgs bosons. 
This is confirmed by the study of simple examples \cite{Maniatis:2014oza}.
For the case that we have
\begin{equation} \label{7.30}
{\cal E}_{H-} = {\cal E}_{A-} = 0
\end{equation}
the field $h_0(x)$ is a mass eigenfield with
mass squared value, see \eqref{7.24},
\begin{equation} \label{7.31}
m_{h_0}^2 = {\cal E}_{--} = - \frac{8}{v_0^2} V^{(0)}.
\end{equation}
In this case the field $h_0(x)$ is called {\em aligned} with
the vacuum expectation value.

\section{Conclusion}

The $n$-Higgs-doublet model has been studied as
a generalization of the THDM and the 3HDM. 
Stability, electroweak symmetry breaking, and the stationary
points
of the Higgs potential have been discussed.
We have presented explicit sets of equations
allowing to determine the stability of any nHDM.
In case of stability, the equations to determine
the stationary points of the potential have been presented.

Of course, there are three types of vacuum solutions:
with complete breaking, with no breaking, and with
partial breaking of \eweakgroup.
For the latter case -- the only one of physical interest --
we have investigated the potential after symmetry breaking.
The mass squared of the physical Higgs bosons have been
given explicitly. For all these
investigations we have found the use of the gauge-invariant 
bilinears very convenient. 
For numerical investigations of the stability and stationarity
equations one has to solve polynomial equations.
For this
there are approaches available, like 
Groebner-bases or homotopy continuation, which
are capable to solve these sets of equations.
We have found that the degree of these 
polynomial equations is independent of the
number $n$ of Higgs bosons;
see \eqref{4.15} to \eqref{4.17}, \eqref{4.20}, \eqref{4.21},
\eqref{6.3} to \eqref{6.7}, and \eqref{7.1} to \eqref{7.3}.
But the number of variables increases, in essence proportional
to $n$.
To conclude: we find it remarkable that, using the method
of bilinears, one can get a rather good overview of the properties
of the potentials of the nHDM, even if at first sight these potentials
seem to be extremely involved.


\begin{acknowledgments}
The work of M.M. was supported, in part, by Fondecyt (Chile)
Grant No. 1140568.
\end{acknowledgments}

\appendix

\section{Generalised Gell-Mann matrices and basis transformations}
\label{appendixA}
Firstly, let us present a construction of the generalised Gell-Mann
matrices $\lambda_a$ of dimension~$n$, that is, $a=1,\ldots,n^2-1$.
We start with defining the $n \times n$ matrix $\tvec{e}_j \tvec{e}_k^\dagger$ with a 1 in the $j$th
row and $k$th column and 0 elsewhere. 
Here $\tvec{e}_j$, $j=1,\ldots,n$, are the Cartesian unit vectors in $\mathbb{C}_n$
\begin{equation} \label{A0}
\begin{split}
&\tvec{e}_1 = \begin{pmatrix} 1,& 0,& \ldots,& 0 \end{pmatrix}^\trans,\\
& \quad \vdots\\
&\tvec{e}_n = \begin{pmatrix} 0,& \ldots,& 0,& 1 \end{pmatrix}^\trans,\\
\end{split}
\end{equation}
In terms of these matrices we construct 
$n^2-1$ hermitian traceless matrices $\lambda_a, a=1,\ldots,n^2-1$ as follows.
With $k=1,\ldots,n-1$ and $j=1,\ldots,k$ we set
\begin{align} 
\label{A.1}
&\lambda_a = \tvec{e}_j \tvec{e}_{k+1}^\dagger + \tvec{e}_{k+1} \tvec{e}_j^\dagger, \quad\text{for } a=k^2+2j-2,\\
\label{A.2}
&\lambda_a = -i \tvec{e}_j \tvec{e}_{k+1}^\dagger + i \tvec{e}_{k+1} \tvec{e}_j^\dagger, \quad\text{for } a=k^2+2j-1.\\
\intertext{In addition we construct $n-1$ diagonal matrices}
\label{A.3}
&\lambda_{(l+1)^2-1} = \sqrt{\frac{2}{l(l+1)}}
\left[ \left( \sum_{j=1}^{l}  \tvec{e}_j \tvec{e}_j^\dagger \right) - l\cdot \tvec{e}_{l+1} \tvec{e}_{l+1}^\dagger \right],\\ 
&\qquad 1 \le l \le n-1. \nonumber
\end{align}
\begin{figure}[t!]
\begin{small}
\begin{tabular}{| >{\centering}m{0.15\columnwidth} | >{\centering}m{0.15\columnwidth} | >{\centering}m{0.15\columnwidth} | >{\centering}m{0.15\columnwidth} | >{\centering}m{0.15\columnwidth} | >{\centering}m{0.15\columnwidth} |}
\hline
0 & $\begin{matrix}1\\ 2\end{matrix}$ & $\begin{matrix}4\\ 5\end{matrix}$ & $\begin{matrix}9\\ 10 \end{matrix}$ & $\cdots$ & $\begin{matrix}(n\!-\!1)^2 \phantom{\!{\tiny +}\!0}\\ (n\!-\!1)^2\!{\tiny +}\!1 \end{matrix}$\\
\tabularnewline
\hline
 & 3  & $\begin{matrix}6\\ 7\end{matrix}$ & $\begin{matrix}11\\ 12\end{matrix}$ & $\cdots$ & $\begin{matrix}(n\!-\!1)^2\!{\tiny +}\!2\\ (n\!-\!1)^2\!{\tiny +}\!3 \end{matrix}$ \\
\tabularnewline
\hline 
 &  & 8  & $\begin{matrix}13\\ 14\end{matrix}$ & $\cdots$ & $\small{\begin{matrix}(n\!-\!1)^2\!{\tiny +}\!4\\ (n\!-\!1)^2\!{\tiny +}\!5\end{matrix}}$\\
\tabularnewline
\hline 
 &  &  & 15  &  $\cdots$ & $\begin{matrix}(n\!-\!1)^2\!{\tiny +}\!6\\ (n\!-\!1)^2\!{\tiny +}\!7\end{matrix}$\\
\tabularnewline
\hline
$\vdots$ & $\vdots$ & $\vdots$ & $\vdots$ & $\cdots$ & $\vdots$ \\
\tabularnewline
\hline 
 & & & & $\cdots$ & $\begin{matrix}  \\ n^2-1 \end{matrix}$\\
\tabularnewline
\hline 
\end{tabular}
\end{small}
\caption{\label{fig-numbering}Numbering scheme for the generalised Gell-Mann matrices
$\lambda_\alpha$ ($\alpha=0, \ldots, n^2-1$).}
\end{figure}
Eventually,  we define
the matrix~$\lambda_0$, proportional to the unit matrix, 
\begin{equation} \label{A.3a}
\lambda_0 = \sqrt{\frac{2}{n}} \unitmatrix_n .
\end{equation}
Let us note that the matrices $\lambda_\alpha$, ($\alpha=0, \ldots, n^2-1$) defined in this way
in particular fulfill the conditions~\eqref{2.4c}.
An easy way to remember this numbering scheme
is as follows. We draw an $n \times n$ square
lattice and insert the numbers $\alpha = 0,1,\ldots,n^2-1$
as shown in Fig.~\ref{fig-numbering}.
If~$\alpha$ is the upper (lower) number in an off-diagonal
square then $\lambda_\alpha$ gets a 1 ($-i$) in this place,
1 ($+i$) in the transposed place, and zero elsewhere.
If $\alpha$ is in a diagonal square $\lambda_\alpha$ is given 
by \eqref{A.3} for $\alpha>0$ and
by \eqref{A.3a} for $\alpha = 0$.

For $n=3$ the matrices $\lambda_a$ ($a=1,\ldots,8$) as constructed above
are the standard Gell-Mann matrices; see for instance \cite{Nachtmann}.

Returning to the case of general $n$ we find it convenient to define also
\begin{equation} \label{A4a}
\begin{split}
\lambda_+ &= \sqrt{\frac{n-1}{n}} \lambda_0 + \sqrt{\frac{1}{n}} \lambda_{n^2-1} =
\sqrt{\frac{2}{n-1}} 
\begin{pmatrix} \unitmatrix_{n-1} & 0 \\ 0 & 0 \end{pmatrix},\\
\lambda_- &= \sqrt{\frac{1}{n}} \lambda_0 - \sqrt{\frac{n-1}{n}} \lambda_{n^2-1} =
\sqrt{2} 
\begin{pmatrix} 0_{n-1} & 0 \\ 0 & 1 \end{pmatrix}.
\end{split}
\end{equation}
The change form the basis $\lambda_0, \lambda_1, \ldots, \lambda_{n^2-2}, \lambda_{n^2-1}$
to $\lambda_+, \lambda_1, \ldots, \lambda_{n^2-2}, \lambda_-$ is made with help of
the following orthogonal $n \times n$ matrix
\begin{equation} \label{A4b}
\begin{split}
 S &= \left(S_{\rho \alpha}\right) =
  \begin{pmatrix}
 S_{+0} & 0 & S_{+,n^2-1}\\
 0 & \unitmatrix_{n^2-2} & 0 \\
 S_{-0} & 0 & S_{-,n^2-1}
 \end{pmatrix} \\
&=
 \begin{pmatrix}
 \sqrt{\frac{n-1}{n}} & 0 & \sqrt{\frac{1}{n}}\\
 0 & \unitmatrix_{n^2-2} & 0 \\
 \sqrt{\frac{1}{n}} & 0 & -\sqrt{\frac{n-1}{n}}
 \end{pmatrix}.
\end{split}
\end{equation}
We have with $\alpha, \beta \in \{0,\ldots, n^2-1  \}$,
\newline
$\rho, \sigma \in \{+,1,\ldots,n^2-2,-\}$
\begin{equation} \label{A4c}
\begin{split}
&S^\trans S = S S^\trans  = \unitmatrix_{n^2}, \qquad S = S^\trans,\\
&\lambda_\rho = S_{\rho \alpha} \lambda_\alpha,\\
&\trace( \lambda_\rho \lambda_\sigma ) = 2 \delta_{\rho \sigma}.
\end{split}
\end{equation}
With the help of $S$ we transform also $K_\alpha$, $\xi_\alpha$, $\tilde{E}_{\alpha \beta}$
(see \eqref{2.6}, \eqref{3.2}, \eqref{pot4}) to the basis 
$\rho, \sigma \in \{+,1,\ldots,n^2-2,-\}$
\begin{equation} \label{A4d}
\begin{split}
&K_\rho = S_{\rho \alpha} K_\alpha = \trace(\twomat{K} \lambda_\rho),\\
&\xi_\rho = S_{\rho \alpha} \xi_\alpha,\\
&\tilde{E}_{\rho \sigma} = S_{\rho \alpha} \tilde{E}_{\alpha \beta} S^\trans_{\beta \sigma}.
\end{split}
\end{equation}
This gives, for instance, with \eqref{eq-kmat}
\begin{equation} \label{A4f}
\begin{split}
K_+ &= \sqrt{\frac{n-1}{n}} K_0 + \sqrt{\frac{1}{n}} K_{n^2-1}\\ 
& =
\sqrt{\frac{2}{n-1}} \big( \varphi_1^\dagger \varphi_1 + \ldots + \varphi_{n-1}^\dagger \varphi_{n-1} \big),\\
K_- &= \sqrt{\frac{1}{n}} K_0 - \sqrt{\frac{n-1}{n}} K_{n^2-1} =
\sqrt{2}  \varphi_n^\dagger \varphi_n.
\end{split}
\end{equation}

\section{Symmetric sums}
\label{appendixB}
Here we want prove the recursive relation~\eqref{Newton} for
the symmetric sums as originally defined in \eqref{defek}.
Consider~$1 \le k \le n$.
First we note that $s_k(\kappa_1, \ldots, \kappa_n)$
is a homogenous function of degree~$k$ in $\kappa_1, \ldots, \kappa_n$.
Therefore we have 
\begin{equation} \label{A.5}
\sum_{l=1}^n \kappa_l \frac{\partial}{\partial \kappa_l} s_k(\kappa_1, \ldots, \kappa_n)
=
k s_k(\kappa_1, \ldots, \kappa_n).
\end{equation}
On the other hand we have
\begin{equation} \label{A.6}
\begin{split}
& \frac{\partial}{\partial \kappa_l} s_k(\kappa_1, \ldots, \kappa_n) = 
\sum_{\stackrel{1 \le i_1 < \ldots < i_{k-1} \le n}{i_r \neq l}} \kappa_{i_1} \cdot \ldots \cdot \kappa_{i_{k-1}} \\
& =  s_{k-1}(\kappa_1, \ldots, \kappa_n) - 
\big(\!\!\!\!\!\!\!\!\!\! \sum_{\stackrel{1 \le i_1 < \ldots < i_{k-2} \le n}{i_r \neq l}} \!\!\!\!\!\! \kappa_{i_1} \cdot \ldots \cdot \kappa_{i_{k-2}} \big) \kappa_l\\
& = 
 s_{k-1}(\kappa_1, \ldots, \kappa_n) -
 s_{k-2}(\kappa_1, \ldots, \kappa_n) \kappa_l\\
& \qquad + \ldots + (-1)^{k-1} s_0 \kappa_l^{k-1}.
\end{split}
\end{equation}
Multiplying in~\eqref{A.6} with $\kappa_l$, summing over $l$, and using~\eqref{A.5},
we get
\begin{multline} \label{A.7}
k s_k(\kappa_1, \ldots, \kappa_n) =
s_{k-1}(\kappa_1, \ldots, \kappa_n)(\kappa_1+ \ldots + \kappa_n)\\
- s_{k-2}(\kappa_1, \ldots, \kappa_n)(\kappa_1^2+ \ldots + \kappa_n^2)\\
 + \ldots + (-1)^{k-1} s_0 (\kappa_1^k+ \ldots + \kappa_n^k)
\end{multline}
which proves~\eqref{Newton}.


\end{document}